\input harvmac
\def\half{{1 \over 2}}
\def\a {{\alpha}}
\def\p {{\partial}}
\def\b {{\beta}}
\def\s {{\sigma}}

\def\ad {{\dot\alpha}}
\def\bd {{\dot\beta}}

\def\t {{\theta}}

\Title{\vbox{\hbox{IFT-P.004/99 }}}
{\vbox{
\centerline{\bf Superspace Action for the First Massive States}
\centerline{\bf of the Superstring}
}}

\bigskip\centerline{Nathan Berkovits\foot{e-mail: nberkovi@ift.unesp.br}}
\bigskip\centerline{Marcelo M. Leite\foot{e-mail: mmleite@ift.unesp.br}}
\bigskip
\centerline{\it Instituto de F\'\i sica Te\'orica, Universidade Estadual
Paulista} 
\centerline{\it Rua Pamplona 145, 01405-900, S\~ao Paulo, SP, Brasil}

\vskip .3in
Using the manifestly spacetime-supersymmetric version of open superstring
field theory, we construct the free action for the first massive states
of the open superstring compactified to four dimensions. This action
is in N=1 D=4 superspace and describes a massive spin-2 multiplet coupled
to two massive scalar multiplets. 

\Date {December 1998}

\newsec{Introduction}

The conventional action for open superstring field theory contains 
contact-term problems caused by the presence of
picture-changing operators\ref\WW{E. Witten,  
Nucl. Phys{\bf B}276(1986) 291\semi
C. Wendt,  
Nucl. Phys{\bf B}314(1989) 209.}. Recently, a new
action was proposed for open superstring field theory where these
picture-changing operators are absent\ref\me{
N. Berkovits, Nucl. Phys{\bf B}450(1995) 90, hep-th/9503099.}. 
In addition to eliminating the
contact-term problems, this new action can be written in a form which
is manifestly N=1 D=4 spacetime-supersymmetric. 

After compactifying
the open superstring on a Calabi-Yau manifold to four dimensions, 
it is easy to show that
the massless compactification-independent contribution to
the free part of this action reproduces the usual N=1 D=4
superspace action for super-Maxwell.   
In this paper, the first massive contribution to the free
part of this action will be computed in N=1 D=4 superspace.

In an earlier paper\ref\Leite{
N. Berkovits and M. M. Leite, 
Phys.Lett.{\bf B}415(1997) 144, hep-th/9709148.}, 
we used open superstring vertex-operator arguments
to show that the massive D=4 spin-two multiplet can
be described in N=1 D=4 superspace by a vector superfield $V_m$
satisfying the constraint $D^\a \s^m_{\a\ad} V_m =(\p^n \p_n +2) V_m =0$.
It will be shown here that
these constraints come from gauge-fixing the equations of motion
of a superspace action involving not just $V_m$, but
also a spinor and two scalar superfields, $V_\a$, $B$ and $C$. The propagating
degrees of freedom of this action 
describe a massive spin-2 multiplet as well as two massive
scalar multiplets which are always present in Calabi-Yau
compactifications of the open superstring to four dimensions.   

Before constructing the open superstring field
theory action for these massive states,
we shall review the construction of the action for
the massless states.

\newsec{Construction of the Free Action for the Massless States}

\noindent

The topological description of the superstring, 
developed by one of the authors with Vafa in 
reference 
\ref\Vafa{
N. Berkovits and C. Vafa, Nucl. Phys{\bf B}433(1995) 123; 
see also N. Berkovits, ``A New Description of the Superstring'', 
proceedings of the VIII J. A. Swieca Summer School on Particles and Fields, 
World Scientific Publishing (1996), hep-th/9604123; Nucl. Phys.{\bf B}431(1994) 
258, hep-th/9404162.},
is particularly suitable for formulating 
open superstring
field theory\me.
In this approach, the BRST current is 
replaced by a spin-one generator $G^{+}$, the
$\eta$ ghost (which comes from bosonizing the RNS super-reparameterization
ghosts as $\beta=\p\xi e^{-\phi}$ and $\gamma=\eta e^\phi$) 
is replaced by
a spin-one generator $\tilde G^+$, and the ghost number current
is replaced by a spin-one generator $J$. As discussed in \Vafa,
these three generators form
part of a `small' twisted N=4 algebra.

The usual condition for physical vertex operators is  
$Q (V)=0$ where $V$ is independent of the $\xi$ zero mode, i.e.
$\eta (V)=0$. Note that $G(V)$ always means the contour integral
of $G$ around $V$.
Since the $\eta$ cohomology is trivial, one
can always find an operator $\Phi$ such that $V=\eta (\Phi)$.
So the condition for $\Phi$ to be physical is that 
$\tilde G^{+}(G^{+}(\Phi)) = 0$  where $\Phi$ is defined up to the
gauge invariance 
$\delta\Phi=G^{+}(\Lambda) + \tilde{G}^{+}(\bar{\Lambda})$. 
These linearized equations of motion and
gauge invariances are easily obtained from the string field theory
action 
\eqn\sftac{S= <\Phi ~G^+ (\tilde G^+ (\Phi))>}
where $<>$ is the two-point
correlation function on a sphere.
As 
usual in open superstrings, the states carry Chan-Paton factors which 
will be suppressed throughout the paper.

The manifestly spacetime-supersymmetric description of the superstring
is related to the usual RNS description by a field redefinition. For
compactifications of the superstring on a Calabi-Yau
threefold, this field redefinition allows N=1 D=4 super-Poincar\'e
invariance to be made manifest.
The field redefinition takes the left-moving
worldsheet matter and ghost fields of
the RNS description into 
five free bosons
$(x^{m},\rho)$ (where $m=0$ to 3) which satisfy the OPE's
\eqn\opes{x^m (y) x^n (z) \to log |y-z| \eta^{mn},\quad
\rho (y) \rho (z) \to log (y-z),} 
eight free fermions 
$(\theta^{\alpha}$, 
${\bar\theta}^{\dot{\alpha}}$, 
$ p^{\alpha}$,
${\bar p}^{\dot{\alpha}})$ (where $\a$ and $\ad$ =1 or 2)
which satisfy the OPE's
$$p_\a (y) \theta^\b (z) \to \delta_\a^\b (y-z)^{-1} ,\quad
\bar p_\ad (y) \theta^\bd (z) \to \delta_\ad^\bd (y-z)^{-1} ,$$
and a 
field theory for the six-dimensional compactification manifold which
is described by the $\hat c=3$ N=2 superconformal generators 
$[T_C,G^+_C,G^-_C,J_C]$. This N=2 superconformal field theory is twisted
so that  
$G^+_C$ has conformal
weight $+1$ and $G^-_C$ has conformal weight $+2$.
As 
usual, right-moving fields are related to the left-moving fields through 
boundary conditions. 

The `small' N=4 superconformal generators are 
defined in
terms of these free fields by
$$T= \half \p x^m \p x_m +p_\a \p\t^\a + \bar p_\ad \p\bar\t^\ad +
\half \p\rho\p\rho -{i\over 2} \p^2\rho + T_C$$ 
$$G^{+} = e^{i\rho}d^{\alpha}d_{\alpha} + G^{+}_C,\quad 
G^{-} = e^{-i\rho}\bar d^{\ad}\bar d_{\ad} + G^{-}_C,$$
\eqn\gt{
\tilde{G}^{+} = e^{-2i\rho+iH_C}\bar d^{\dot\alpha}\bar d_{\dot\alpha} 
+ e^{-i\rho +i H_c} ({G}^{-}_C),}
$$\tilde{G}^{-} = e^{2i\rho-iH_C} d^{\alpha} d_{\alpha} 
+ e^{i\rho -i H_c} ({G}^{+}_C), $$
$$J=i\p\rho +i J_C,\quad J^{++}=e^{-i\rho +i H_C},\quad 
J^{--}=e^{+i\rho -i H_C}, $$
where 
\eqn\ddef{d_{\alpha} = p_{\alpha}
+ {i\over 2} \bar\theta^{\dot\alpha}\partial x_{\alpha \dot{\alpha}}
- {1\over 4}(\bar\theta)^{2}\partial\theta_{\alpha}
+{1\over 8}\theta_{\alpha}\partial(\bar\theta)^{2}
,}
$$\bar{d_{\dot\alpha}} = \bar p_{\dot\alpha}
+ {i\over 2}\theta^{\alpha}\partial x_{\alpha \dot{\alpha}}
- {1\over 4}(\theta)^{2}\partial\bar\theta_{\dot\alpha}
+{1\over 8}\bar\theta_{\dot\alpha}\partial(\theta)^{2},$$
$J_C=\p H_C, $ 
and we use the bispinor convention $v_{\a\ad}=\sigma^m_{\a\ad} v_m$. 
Our Pauli matrices, $\sigma^m_{\a\ad}$ and $\bar\sigma_m^{\ad\a}$, are
those of Wess and Bagger\ref\WB{J. Wess and J. Bagger,
{\it Supersymmetry and Supergravity: Notes
from Lectures given at Princeton University}, Princeton Univ. Press, 1982}, 
but we
choose to 
define $(\t)^2=(\t^\a\t_\a)$ and 
$(\bar\t)^2=(\bar\t^\ad\bar\t_\ad)$, and to  
use the `mostly-negative' Minkowski metric
$\eta^{mn}= diag(+1,-1,-1,-1)$ so that $Tr(\sigma^m\bar\sigma^n)=2\eta^{mn}$.

Although our hermiticity conditions
on the superspace variables are $(x^m)^*=x^m$ and $(\t^\a)^*=\bar\t^\ad$ 
as usual, our hermiticity conditions on the chiral bosons are  
$$\rho^* = 2\rho- H_C,\quad 
H_C^* = 3\rho- 2 H_C, $$ 
so that $(G^+)^* = \tilde G^+$.  
Note that $d_\a$ and $\bar d_\ad$ have been defined to commute
with the spacetime supersymmetry generators and to satisfy the 
OPE's
\eqn\ddope{ 
d_{\alpha}(y) \bar d_{\dot\alpha}(z)\to i{\Pi_{\alpha
\dot{\alpha}}\over{y-z}}, \quad 
d_{\alpha}(y)\Pi_{\beta \dot{\beta}}(z)\to
{{-2i\epsilon_{\alpha\beta}\partial \bar\theta_{\dot{\beta}}}\over{y-z}}} 
where 
$$\Pi_{\alpha \dot{\alpha}}= \partial x_{{\alpha \dot{\alpha}}}
+ i\theta_\alpha\partial{\bar\theta_{\dot\alpha}}
+ i\bar\theta_{\dot\alpha}\partial\theta_{\alpha}.$$

To make N=1 D=4 supersymmetry manifest in the string
field theory action, one integrates over the non-zero modes
of the worldsheet fields in the correlation function of \sftac, 
but leaves the integration over the 
zero modes of $x^m$,
$\t^\a$, and $\bar\t^\ad$ explicit in the action. The 
result is the N=1 D=4 superspace action  
\eqn\super{S=\int d^4 x d^2 \t d^2 \bar \t   ~<\Phi G^+ (\tilde G^+ (\Phi))>}
where $<>$ now does not include the contribution of the zero modes
of $x^m$, $\t^\a$, and $\bar\t^\ad$. 

As our first example, we shall consider the contribution to the above
action from the 
massless compactification-independent
states. Since there is no tachyon, the massless states at zero
momentum are described
by string fields of conformal
weight zero. The string field is independent of the compactification
manifold and is of neutral U(1)-charge (i.e. zero ghost number), so it is 
described by a scalar superfield 
$V(x,\t,\bar\t)$. Plugging $\Phi=V$ into the action of 
\super, one reproduces
the super-Maxwell action
\eqn\max{S=\int d^{4}x d^{2}\theta d^{2}\bar\theta 
V{\bar D}_{\dot\alpha}D^{2}{\bar D}^{\dot\alpha}V} 
where $D_\a = {\p \over {\p\t^\a}}+{i\over 2} 
\bar\t^\ad \sigma^m_{\a\ad} \p_m$
and 
$\bar D_\ad = {\p \over {\p\bar\t^\ad}}+
{i\over 2} \t^\a \sigma^m_{\a\ad} \p_m$.
Furthermore, the gauge invariances 
$\delta\Phi=G^{+}(\Lambda) + \tilde{G}^{+}(\bar{\Lambda})$
reproduce the expected gauge invariances 
$$\delta V = D^2 \lambda + \bar D^2 \bar\lambda$$
where $\Lambda = e^{-i\rho}\lambda$ and $\bar\Lambda = e^{2i\rho-i H_C}
\bar\lambda$. 

\newsec{Action for First Massive States}

The field theory action of \super\ will now be used to compute 
the contribution
of the first massive states of the open superstring. We
shall consider a generic Calabi-Yau three-fold and will not
consider states which depend on the explicit structure of
the compactification manifold. However, we will allow states
which can be constructed from the Calabi-Yau U(1) current $\p H_C$.
As will be shown later, the presence of
such states in the action is necessary for
gauge invariance.

Since we want to describe states of $(mass)^2=2$, our string field should
contain conformal weight $+1$ at zero momentum. Furthermore, 
it should have no U(1) charge and should only depend on the compactification
manifold through the U(1) current. The most general
such string field is 
$$\Phi = d^{\alpha}W_{\alpha}(x,\theta,\bar\theta) 
+ \bar d^{\dot\alpha}\bar W_{\dot\alpha}(x,\theta,\bar\theta) 
+ \Pi^{m}V_m(x,\theta,\bar\theta)
+\partial\theta^{\alpha}V_{\alpha}(x,\theta,\bar\theta)
+\partial\bar\theta^{\dot\alpha}\bar V_{\dot\alpha}(x,\theta,\bar\theta)$$
\eqn\sfield{+i(\p\rho -\partial H_C ) ~B(x,\theta,\bar\theta)
+(\partial H_C -3\p\rho)  ~C(x,\t,\bar\t)} 
where $B$ and $C$ are real superfields. 

This field 
transforms as $\delta\Phi = G^+(\Lambda)+\tilde G^+(\bar\Lambda)$
under the gauge transformations parameterized by 
$$\Lambda = e^{-i\rho}
(d^{\alpha}C_{\alpha} + \bar{d^{\dot\alpha}}\bar{E_{\dot\alpha}} 
+ \Pi^{m}B_{m} + \partial\rho F + 
+\partial\theta^{\alpha}B_{\alpha} + 
\partial{\bar\theta}^{\dot\alpha}{\bar H}_{\dot\alpha}), $$
$$\bar\Lambda = 
e^{2i\rho - iH_{C}}
(d^{\alpha}E_{\alpha} + \bar{d^{\dot\alpha}}\bar{C_{\dot\alpha}} 
+ \Pi^{m}\bar B_{m} + (2\partial\rho-\partial H_C) \bar F 
+\partial\theta^{\alpha}H_{\alpha} + 
\partial{\bar\theta}^{\dot\alpha}{\bar B}_{\dot\alpha}). $$
Note that $\delta \Phi=0$ when $\Lambda = G^+ \Omega$ or
$\bar\Lambda = \tilde G^+ \bar\Omega$. Defining 
$$\Omega = e^{-2i\rho}(\p^2 \t^\a M_\a + (\p\t)^2 N + \p\t^\a \bar d^{\ad}
P_{\a\ad} + \Pi_{\a\ad}\p\t^\a \bar Q^{\ad}),$$
one can see that the transformations parameterized by
$C_\a$, $B_m$, $F$, and $H_\a$ 
can be ignored. Furthermore, the transformations parameterized
by $B_\a$ and $\bar B_\ad$ can be used to algebraically gauge-fix
$W_\a=\bar W_\ad=0$ in the string field $\Phi$. 

In this gauge, the remaining string fields transform under the
remaining gauge transformations as:
\eqn\rem{\delta V_{m} = 
- 2i (\sigma_{m})_{\alpha \dot{\alpha}}D^{\alpha}\bar{E^{\dot\alpha}} 
- 2i (\sigma_{m})_{\alpha \dot{\alpha}}\bar D^{\dot\alpha}E^{\alpha},} 
$$\delta V_{\alpha} =  4E_{\alpha} - 
 {1\over 2}D^{2}\bar D^{2}E_{\alpha}  
+ i\p_{\alpha \dot\alpha} D^{2}\bar E^{\dot\alpha},$$
$$\delta (C+iB) = i D^{\alpha}\bar D^{2} E_{\alpha}.$$
Note that $B$ and $C$ are not gauge-invariant, so they can not be dropped
from the action without breaking gauge invariance. In our earlier
analysis of massive vertex operators \Leite, it was not necessary
to include $B$ and $C$ since we were working in a fixed gauge. 

Finally, the string field action 
$S=\int d^4 x d^2 \t d^2 \bar \t   ~<\Phi G^+ (\tilde G^+ (\Phi))>$
in the gauge $W_\a=\bar W_\ad =0$ is given by 
\eqn\action{S= \int d^4 x d^2 \theta d^2 \bar\theta  [
-V^{m} ({\bar D}^{2}D^{2}V_{m} - 4V_{m} 
+ 2i(\bar\sigma^n)^{\dot{\alpha} \alpha}\partial_{n}
{\bar D}_{\dot\alpha}D_{\alpha}V_{m}  } 
$$
-4i(\bar\sigma_m)^{\dot{\alpha} \alpha}(D_{\alpha}\bar V_{\dot\alpha} 
+{\bar D}_{\dot\alpha}
V_{\alpha})
- 8 \partial_m B + 12(\bar\sigma_m)^{\dot{\alpha} \alpha}
[{\bar D}_{\dot\alpha},D_{\alpha}]C) $$
$$
+ V^{\alpha} ( - 
4{\bar D}^{\dot{\alpha}}D_{\alpha}\bar V_{\dot\alpha} - 
2{\bar D}^{2}V_{\alpha} 
+ 24\p_{\alpha \dot{\alpha}}{\bar D}^{\dot\alpha}C
- D_{\alpha}{\bar D}^2(18iC +2B) )
$$
$$+ \bar V^{\dot\alpha} (
-2 D^{2}\bar V_{\dot\alpha}
-24\p_{\alpha \dot{\alpha}} D^{\alpha}C
+{\bar D}_{\dot\alpha}D^{2}(18iC -2B) )$$
$$
-3C(11 \bar D^2 D^2 -16\p^m \p_m  -8)C
+B
(- \bar D^2 D^2 -8)B +6B \p_{\a\ad}[\bar D^\ad, D^\a]C~].$$

To evaluate the propagating degrees of freedom, it is convenient
to use the gauge transformations parameterized by $E_\a$ and $\bar E_\ad$
to gauge-fix $V_\a =\bar V_\ad=0$. Note that this gauge-fixing involves
derivatives, so unlike the gauge-fixing of $W_\a$, it cannot be performed
directly in the action. Furthermore, $\delta
V_\a=0$ when the gauge parameter $E_\a$ satisfies 
$E_\a ={i\over 2}\p_{\a\ad}D^2 \bar E^\ad $ and $(\p^n \p_n -1) E_\a =0$. 
This implies that there is a residue
gauge transformation even after gauge-fixing $V_\a=0$. 

In the gauge $V_\a=\bar V_\ad=0$, the equations of motion from the
action of \action\ are 
\eqn\vv{{1\over 2}\{ \bar D^{2},D^{2}\}V_{m} - 4V_{m} -2\p^n\p_n V_m
- 4 \partial_m B + 6(\bar\sigma_m)^{\dot{\alpha} \alpha}
[{\bar D}_{\dot\alpha},D_{\alpha}]C = 0,}
\eqn\valpha{
4i(\sigma^{m})_{\alpha \dot{\alpha}}{\bar D}^{\dot\alpha}V_{m} 
+ 24\p_{\alpha \dot{\alpha}}{\bar D}^{\dot\alpha}C
+ D_{\alpha}\bar D^2 (-18iC-2B)=0 ,}
\eqn\bb{
-8\partial^{m}V_{m} -16B -\{\bar D^2, D^2\}B + 6 \p_{\a\ad}[\bar D^\ad,
D^\a] C =0,}
\eqn\cc{-12 \bar\sigma_m^{\ad\a}[\bar D_\ad,D_\a] V^m +48C
-33\{\bar D^2, D^2\}C +96\p_m\p^m C - 6 \p_{\a\ad}[\bar D^\ad,
D^\a ] B =0.}

Plugging \valpha\ into \bb\ and \cc, one learns that
\eqn\solution
{B={i\over 8} [D^2,\bar D^2] C}
\eqn\motion
{16 (\p^m\p_m -1)C = 6 \{D^2, \bar D^2\} C.}
The equation of motion for $C$ has two solutions:
\eqn\sola{a) ~~~ D^2 C =0,\quad (\p^m \p_m -1)C=0;}
\eqn\solb{b) ~~~\bar D^\ad D^2 \bar D_\ad C =0,\quad (\p^m \p_m +2)C=0.}

Using the residue gauge transformations of $E_\a$, 
any solution of equation \sola\ can be gauged away. Therefore,
the only physical degrees of freedom of $C$ come from solutions of
\solb, which are of the form:
\eqn\prop{C = F + \bar F }
where $F$ is a chiral superfield satisfying 
$$\bar D_\ad F = D_\a \bar F = (\p^m \p_m +2) F = (\p^m \p_m +2)\bar F =0.$$
Although $F$ is a single chiral superfield, it
describes two massive scalar multiplets since it does
not satisfy the on-shell equation $D^2 F =0$. Expanding in components,
$$F= X +\t^\a \xi_\a + (\t)^2 Y$$
where $X$ and $Y$ are complex bosons
satisfying $(\p^m \p_m +2) X=
(\p^m \p_m +2) Y= 0$. Defining $\psi_\ad =-{i\over{\sqrt{2}}}\p_{\a\ad}\xi^\a$, 
the equations of motion for the fermion are
$$\p_{\a\ad} \xi^\a= i\sqrt{2} \psi_{\ad},\quad 
\p_{\a\ad} \psi^\ad= i\sqrt{2} \xi_\a $$
which describe a massive Dirac spinor $\Lambda^A =(\xi^\a,\psi^\ad)$.

Plugging \prop\ and \solution\ into \vv\ and \valpha, one obtains
\eqn\solv{D^\a V_{\a\ad} -2i \p_{\a\ad} D^\a F=0,}
\eqn\solvvv{{1\over 2}\{\bar D^{2},D^{2}\}V_{m} - 4V_{m} -2\p^n\p_n V_m
-16i (F-\bar F)=0.}
Finally, shifting $\hat V_m = V_m -2i\p_m (F-\bar F)$, 
one can rewrite \solv\ and \solvvv\ as 
\eqn\spintwo{D^\a \hat V_{\a\ad} =0,\quad (\p^m\p_m +2)\hat V_m=0,}
which were shown in
\Leite\ to be the gauge-fixed equations of motion of a massive spin-two
multiplet. Therefore, the propagating degrees of freedom described by the
action of \action\ are a massive spin-two multiplet and two massive scalar
multiplets. 

On-shell, these multiplets contain 12 bosonic and 12 fermionic
degrees of freedom. The bosonic degrees of freedom can be 
identified with the following 12 
Neveu-Schwarz light-cone gauge vertex operators:
\eqn\lc{b_{-{3\over 2}}^m |0\rangle,\quad 
b_{-{1\over 2}}^m a_{-1}^n |0\rangle,}
$$\omega_{JKL} b_{-{1\over 2}}^J b_{-{1\over 2}}^K b_{-{1\over 2}}^L 
|0\rangle,\quad 
\bar\omega_{\bar J\bar K\bar L} b_{-{1\over 2}}^{\bar J} 
b_{-{1\over 2}}^{\bar K} b_{-{1\over 2}}^{\bar L} |0\rangle,$$
$$
b_{-{1\over 2}}^m g_{J\bar J} b_{-{1\over 2}}^J 
b_{-{1\over 2}}^{\bar J} |0\rangle,\quad 
g_{J\bar J} b_{-{1\over 2}}^J a_{-1}^{\bar J} |0\rangle,\quad 
g_{J\bar J} b_{-{1\over 2}}^{\bar J} a_{-1}^{J} |0\rangle ,$$
where $m,n,p=2,3$ are the D=4 light-cone indices, 
$g_{J \bar J}$ and $\omega_{JKL}$ are the metric
and holomorphic 3-form used to define the
Calabi-Yau manifold, and $b^m_N$ and $a^m_N$ are 
the oscillator modes of the light-cone $\psi^m$ and $\p X^m$. 
Similarly,
the fermionic degrees of freedom can be identified with the following
12 Ramond light-cone gauge vertex operators: 
\eqn\ferm{ b_{-1}^m |0\rangle^{++++}, \quad b_{-1}^m |0\rangle^{----},\quad 
a_{-1}^m |0\rangle^{-+++}, \quad a_{-1}^m |0\rangle^{+---},}
$$g_{J\bar J} b_{-1}^J b^{\bar J}_0 |0\rangle^{+---},\quad 
g_{J\bar J} a_{-1}^J b^{\bar J}_0 |0\rangle^{----},$$ 
$$g_{J\bar J} b_{0}^J b^{\bar J}_{-1} |0\rangle^{-+++},\quad 
g_{J\bar J} b_{0}^J a^{\bar J}_{-1} |0\rangle^{++++}$$ 
where $|0\rangle^{\pm\pm\pm\pm}$ is the light-cone spinor in
SU(4) notation where the
last 3 $\pm$ signs refer to the six 
Calabi-Yau directions. 
Although
there might be other states 
of $(mass)^2= 2$ in the open superstring spectrum which 
come from Calabi-Yau excitations, 
the 12+12 states described above 
are always present in any Calabi-Yau compactification. 

\vskip 20pt 
\centerline {\bf Acknowledgements}

NB would like to acknowledge partial support from CNPq grant number
300256/94-9 and MML would like to acknowledge support from FAPESP
grant number 96/03546-7.

\listrefs

\end